# Efficient Fastest-Path Computations in Road Maps

Renjie Chen
Max-Planck Institute for Informatics
Saarbrucken, Germany

Craig Gotsman
New Jersey Institute of Technology
Newark, NJ, USA

**Abstract**

In the age of real-time online traffic information and GPS-enabled devices, fastest-path computations between two points in a road network modeled as a directed graph, where each directed edge is weighted by a "travel time" value, are becoming a standard feature of many navigation-related applications. To support this, very efficient computation of these paths in very large road networks is critical. Fastest paths may be computed as minimal-cost paths in a weighted directed graph, but traditional minimal-cost path algorithms based on variants of the classic Dijkstra algorithm do not scale well, as in the worst case they may traverse the entire graph. A common improvement, which can dramatically reduce the number of traversed graph vertices, is the A* algorithm, which requires a good heuristic lower bound on the minimal cost. We introduce a simple, but very effective, heuristic function based on a small number of values assigned to each graph vertex. The values are based on graph separators and computed efficiently in a preprocessing stage. We present experimental results demonstrating that our heuristic provides estimates of the minimal cost which are superior to those of other heuristics. Our experiments show that when used in the A* algorithm, this heuristic can reduce the number of vertices traversed by an order of magnitude compared to other heuristics.

## 1. Introduction

### The Shortest, Minimal-Cost and Fastest Path Problems

The shortest-path problem on graphs is one of the most fundamental algorithms in computer science, the graph being one of the most basic and common discrete structures, modeling an abundance of real-world problems involving networks. In the most basic scenario, graph vertices represent entities in a network and an edge between two vertices indicates the existence of a link between them (e.g. a communication or social network). The shortest path between two vertices $s$ and $t$ in the graph is then the path between $s$ and $t$ containing the minimal number of edges. In the case of a communication network, this could be the cheapest way to route a message to $t$, originating at $s$. In the more general case, edges are assigned weights which measure a cost associated with traversing that edge. The shortest path then becomes a minimal-cost path, where the cost of the path is the sum of the costs of its edges. In the case of a communication network, the associated cost of an edge may be its conductance.

A very important type of network is a road map, the graph vertices representing road junctions and the edges road segments between the junctions. In the simplest scenario, the graph is planar, the vertices are embedded in the plane, namely have $(x, y)$ coordinates, and each edge is assigned a positive weight measuring its Euclidean length in the plane. The minimal-cost path between two vertices $s$ and $t$ is then the edge path of minimal Euclidean length between $s$ and $t$, which could indicate the shortest drive (or walk) between these two points. In practice, in vehicle navigation applications, not all the edges of the road map having the same length are equivalent for a driver, since the possible driving speed on the roads may vary, depending on the category of the road. Highways are usually preferred, as they allow for higher speeds, thus a faster drive. Consequently, the more relevant weight assigned to a road segment is the so-called "travel time", which is the segment length divided by the maximal speed possible on that segment. The resulting minimal-cost path in this weighted graph is sometimes called the "fastest path". The more realistic variant of this problem is when the graph edges are *directed*, namely the travel time along an edge may depend on the direction of the edge. In the special case of a one-way road, the edge exists in just one direction (or, equivalently, the travel time in the opposite direction is infinite).

**Precomputation and Dynamic Fastest Path Problems**

Traditional minimal-cost path algorithms do not scale well to very large networks. More practical algorithms rely on a (typically heavy) *preprocessing* of the graph, resulting in extra information to store along with the basic graph data, which is exploited in answering *online* $(s, t)$-minimal-cost queries efficiently. While effective, this approach introduces a complication. Using travel times as the costs of road network edges is useful to correctly model a real-world navigation problem, but it also imposes a dynamic character on the problem, as the maximal speed on a road segment is rarely a constant – it changes over time depending on traffic conditions – hence the travel times are dynamic. Consequently, an algorithm which relies on preprocessing of the graph data in order to speed up the online $(s, t)$-fastest-path queries, must deal with the dynamic nature of the data by periodic repetition of the preprocessing. This rules out the use of unduly heavy preprocessing.

**Objective**

The computation of minimal-cost paths in dynamic weighted graphs has been the subject of intense study over the past decades, and many techniques have been proposed to solve different variants of the problem. A complete survey of the state-of-the-art is beyond the scope of this paper and we refer the interested reader to the survey and comparison of Bast et al. [1].

Our contribution is the description of a very effective *heuristic* function which can be used in the well-known A* algorithm for minimal-cost path computation on weighted directed graphs. Computation of the heuristic is fast and can easily be repeated periodically to accommodate dynamic traffic conditions in road networks. A* with this heuristic can be used in conjunction with many other techniques to provide a more complete solution to the general problem.

## 2. The Dijkstra and A* Algorithms

While in practice we typically would like to solve the "point-to-point" minimal-cost path problem between a *source* vertex $s$ and a *target* vertex $t$ in a directed graph, it turns out that this indirectly involves computing the minimal-cost path from $s$ to many other vertices in the graph. Let $G = (V, E, w)$ be a directed graph with vertex set $V$ and edge set $E \subset V \times V$, such that the positive real cost of traversing the directed edge $(u, v)$ is $w(u, v)$. The minimal cost of a path from given vertex $s \in V$ to given vertex $t \in V$ may be obtained by computing the entire function $c_s(v)$ from $s$ to any other vertex $v \in V$ by solving the following linear program:

$$\max c_s(t)$$

$$s.t. \quad c_s(s) = 0, \quad \forall edge\ (u, v) \in E(G): \ c_s(v) - c_s(u) \leq w(u, v)$$

Thinking of $c_s(v)$ as an "embedding" of the graph vertices on the real line, this means we would like to "stretch" $s$ and $t$ as far apart as possible on the line, subject to the constraint that the endpoints of any edge $(u,v)$ are separated by a distance of at most $w(u,v)$ - the weight of the edge.

Denote by $\nabla(u,v) = \frac{c_s(v) - c_s(u)}{w(u,v)}$ the "gradient" of $c_s$ along the edge $e$. In the optimal solution $\nabla(u,v) = 1$ for all edges along the minimal-cost path, and $\nabla(u,v) \leq 1$ for all other edges. Thus a gradient-descent path of $c$ starting at $s$ traces out the minimal-cost path.

In practice, this linear program can be transformed into a dynamic programming problem, which in turn can be solved by the celebrated Dijkstra algorithm [4], which traverses the graph vertices guided by a priority queue of vertices. The procedure terminates if $t$ is reached and a minimal-cost path is then generated by tracing the path backwards from $t$. If the priority queue empties before $t$ is reached, the search fails and a minimal-cost path does not exist (e.g. if the graph is not connected). The complexity of the most efficient implementation of the Dijkstra algorithm [5] is $O(m + n \log n)$, where $m$ is the number of graph edges and $n$ the number of graph vertices. Unfortunately, this is prohibitive, for the number of edges $m$ is typically *much* larger than the number of edges along the minimal-cost path.

The Dijkstra algorithm may be accelerated into a "guided" A* search [7] if there is additional domain knowledge in the form of a *heuristic* function $h(v,t)$, one that estimates the minimal cost from $v$ to $t$. The simplest example of a heuristic function for a plane graph with edge-length weights is the planar Euclidean distance $h(v,t) = \|p(v) - p(t)\|_2$, where $p(v)$ are the 2D coordinates of the position of $v$ in the plane. A* is guaranteed to find the shortest path if $h$ is *admissible*, namely is a *lower bound* on the true minimal cost. It is easy to see that the Euclidean distance mentioned above has this property. Like the Dijkstra algorithm, A* maintains a priority queue of *OPEN* vertices. If $h$ is not admissible, a path will still be found, but not necessarily the minimal-cost path.

If $h$ satisfies the additional *consistency* (or *monotonicity*) condition $h(v,t) - h(u,t) \leq w(u,v)$ for every edge $(u,v)$ of the graph and every vertex $t$, then A* can be implemented more efficiently - no node needs to be processed more than once - and A* is equivalent to running Dijkstra's algorithm with the modified (still positive) edge weights: $w'(u,v) = w(u,v) + h(u,t) - h(v,t)$. In practice, in addition to the *OPEN* priority queue, a list *CLOSED* is maintained. Once popped from *OPEN*, a vertex goes into *CLOSED* and is never considered again.

Note that the original Dijkstra algorithm is equivalent to A* with the trivial admissible and consistent heuristic $h(v,t) \equiv 0$.

The following two theorems are useful in characterizing heuristics.

**Theorem 1:** If $h$ is consistent and $h(t,t) = 0$, then $h$ is admissible.

**Proof:** By induction on minimal cost from $t$. ♦

**Theorem 2:** If $h$ is derived from a *metric* function $m$, namely $h(u,t) = m(u,t)$ and $m(u,v) \leq w(u,v)$ for all edges $(u,v)$, then $h$ is consistent.

**Proof:** Apply the triangle inequality and symmetry of $m$: $h(v,t) - h(u,t) = m(v,t) - m(u,t) \leq m(v,u) = m(u,v) \leq w(u,v)$. ♦

Theorems 1 and 2 imply that the planar Euclidean heuristic mentioned above is admissible and consistent. The admissible heuristic $h_1$ is called more *informed* than the admissible heuristic $h_2$ if $h_1(v,t) \geq h_2(v,t)$ for all $v,t \in V$.

## 2.1 A* Heuristics

Much effort has been invested in designing good heuristics for A*. A complete account would be lengthy, and much of it is domain-dependent, so we discuss here just the most generic methods.

**The Optimal Heuristic**

In general, it is possible to precompute the optimal heuristic $h(v,t)$ by solving a convex semi-definite program (SDP) for the $O(n^2)$ values of $h$, forcing the conditions necessary for the heuristic to be admissible and consistent [9]. It relies on the convenient fact that it is sufficient for the heuristic to be "locally" admissible on single edges, namely $h(u,v) \le w(u,v)$ for every edge $(u,v)$ in order that it be admissible over arbitrary paths, significantly reducing the number of linear inequality conditions in the semi-definite program to $O(m)$, thus the complexity of the entire algorithm to $O(m^3)$. However, this complexity is still prohibitive and the method is not applicable to graphs containing more than a few thousand vertices. A number of improvements to this are possible, but the method still remains quite complicated.

**The Differential Heuristic DH**

A very simple, but surprisingly effective *differential heuristic*, was proposed by Goldberg et al. [6] (who called it *ALT*) and independently by Chow [2]. It requires some preprocessing of the graph $G = (V, E, w)$. A small number (usually $k \le 10$) "landmark" vertices (also called anchors/pivots/centers) $l_1, ..., l_k$ are chosen from $V(G)$. In a preprocessing step, for each vertex $v \in V(G)$, the vector of minimal costs $c(v) = \big(c(l_1, v), ..., c(l_k, v)\big)$ is computed and stored. Then, at the online computation of the minimal-cost path from $s$ to $t$, the heuristic

$$h(v,t) = \max_{1 \le i \le k}\{|c(l_i, v) - c(l_i, t)|\} \tag{1}$$

is used. This heuristic requires $O\big(k(m + n \log n)\big)$ preprocessing time and $O(kn)$ space to store. Given $v$ and $t$, $h(v,t)$ can be computed online in $O(k)$ time.

It is convenient to think of $c(v)$ as an *embedding* of $v$ in $R^k$ and $h(v)$ as the embedding distance between $v$ and $t$ using the $l_\infty$ norm:

$$h(v,t) = \|c(v) - c(t)\|_\infty$$

It is easy to see that $h(v,t)$ is exact, namely $h(v,t) = c(v,t)$, if $v$ is on one of the minimal-cost paths between $l_i$ and $t$. It is also easy to apply Theorems 1 and 2 to show that the differential heuristic is admissible and consistent.

The degrees of freedom in this heuristic are the choice of the landmark vertices. Goldberg et al [6] show how to optimize these, concluding that a good choice are landmarks which cover the graph well. In the special case of a plane (or close to plane) graph, a good choice are vertices covering the boundary. In the sequel we will call this heuristic *DH*.

**The FastMap Heuristic FM**

Inspired by the interpretation of the differential heuristic as an embedding distance and by the *FastMap* algorithm used in machine learning, Cohen et al. [3] devised another embedding based on *pairs* of landmarks and defined the heuristic using the $l_1$ norm distance between the embeddings.

The algorithm proceeds by finding a pair of *farthest* vertices $(a_1, b_1)$ – those having a large minimal-cost path between them - and computing for every vertex $v$:

$$f_1(v) = \frac{1}{2}(c(a_1, v) - c(b_1, v)),$$

Defining the following function on pairs of vertices:

$$h_1(u, v) = |f_1(u) - f_1(v)|$$

the weight $w(u, v)$ of the graph edge $(u, v)$ is then modified by subtracting $h_1(u, v)$ from it and the process repeated $k - 1$ times on the modified graph to obtain the embedding vector $r(v) = (f_1(v), \ldots, f_k(v))$. The final heuristic is the $l_1$ embedding distance:

$$h(v, t) = \|r(v) - r(t)\|_1$$

The authors show that this heuristic is also admissible and consistent. In the sequel we will call this heuristic *FM*.

## 3. The Separator Heuristic SH

Since each landmark employed by the differential heuristic defines a *cost field* on the graph vertices, where every vertex is assigned the value of the minimal cost of a path between vertex and the landmark, we first observe that this concept may be easily generalized. Instead of a landmark being a mere *single* vertex, it may be a *set* of vertices $S \subset V(G)$, and the cost of a vertex $v$ (relative to $S$) is defined as:

$$c(v, S) = \min_{u \in S} c(v, u)$$

This defines a more complicated distance field per landmark, to which the triangle inequality may be applied to obtain an analogous differential heuristic. Unfortunately, in practice this generalization does not add much power to that heuristic.

Significantly more power can be obtained if the set $S$ is a *separator* of the graph, namely its removal (along with the edges incident on the removed vertices) results in $V$ being partitioned into three sets $U_1$, $S$ and $U_2 = V - U_1 - S$, such that there exists no edges between $U_1$ and $U_2$. This means that $S$ *separates* between $U_1$ and $U_2$ and the separated graph contains at least two connected components, none of them mixing $U_1$ and $U_2$. We may take advantage of the dichotomy on $V$ induced by $S$ by defining a *signed cost field* on $V$ – positive in $U_1$ and *negative* in $U_2$. Denote this signed cost field by $C$.

Fig. 1 shows the unsigned cost fields induced on a road network by a single landmark vertex or a set of 5 landmark vertices, compared to the signed cost field induced by a separator.

As with the differential heuristic, we choose $k$ separators $S_1, \ldots, S_k$, and define the embedding

$$r(v) = (C(v, S_1), \ldots, C(v, S_k))$$

and the resulting heuristic is the $l_\infty$ embedding distance:

$$h(v, t) = \|r(v) - r(t)\|_\infty$$

Using a signed cost boosts the values of this heuristic significantly. It remains to show that it is still admissible and consistent. Since the use of the signed cost changes the rules of the game relative to the differential heuristic, we provide next a separate proof of admissibility and consistency. In the sequel we will call this heuristic *SH*.

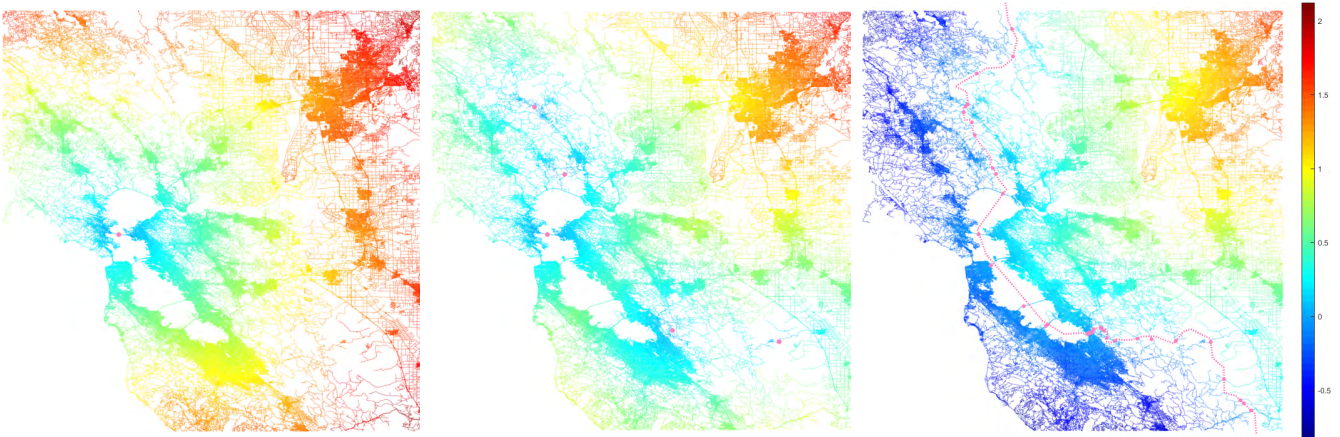

**Figure 1:** Cost fields on undirected graph with edges weighted by Euclidean edge lengths: **(left)** Unsigned cost field induced by a single (magenta) landmark. **(middle)** Unsigned cost field induced by a set of 5 landmarks. **(right)** Signed cost field induced by a separator.

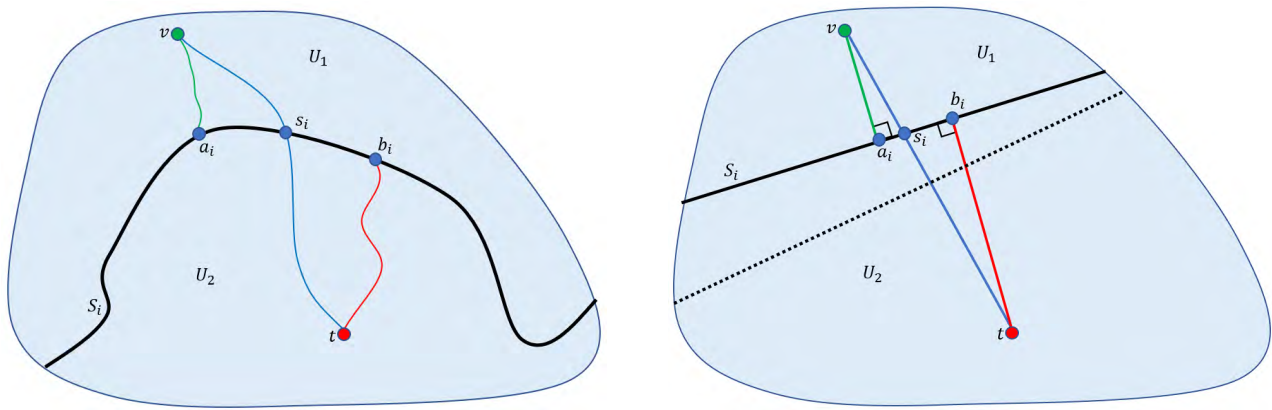

**Figure 2: (left)** Illustration of Case 2 of proof of Theorem 3. The blue path is the minimal-cost path between vertices $v$ and $t$, which must cross the separator $S_i$ at some vertex $s_i$. Vertices $a_i$ and $b_i$ are those on the separator having minimal cost to $v$ and $t$, respectively. **(right)** Analogous scenario for the planar Euclidean distance function. If $S_i$ is approximately parallel to the (dotted) bisector between $v$ and $t$, $c(a_i, b_i)$ will be small and $h(v, t)$ more informed.

**Theorem 3:** The separator heuristic is admissible and consistent.

**Proof:** Assume $S_i$ separates $G$ into $U_1$ and $U_2$. Denote by $a_i$ and $b_i$ the vertices of $S_i$ with minimal cost to $v$ and $t$, respectively.

**Case 1:** $v, t \in U_1$ or $v, t \in U_2$. In this case $C(v, a_i)$ and $C(t, b_i)$ have the same sign and this case is similar to that of the differential heuristic. By definition:

$$c(v, a_i) \leq c(v, b_i) \quad \text{and} \quad c(t, b_i) \leq c(t, a_i)$$

By the triangle inequality:

$$c(v, b_i) \leq c(t, b_i) + c(v, t)$$

so

$$c(v, t) \geq c(v, b_i) - c(t, b_i) \geq c(v, a_i) - c(t, b_i)$$

Again by the triangle inequality:

$$c(t, a_i) \leq c(v, a_i) + c(v, t)$$

so

$$c(v, t) \geq c(t, a_i) - c(v, a_i) \geq c(t, b_i) - c(v, a_i)$$

Putting these two together:

$$c(v, t) \geq |c(v, a_i) - c(t, b_i)| = |C(v, a_i) - C(t, b_i)| = |C(v, S_i) - C(t, S_i)|$$

**Case 2:** $v \in U_1$ and $t \in U_2$ or vice versa. See Fig. 2 (left). Since $S_i$ separate $U_1$ and $U_2$, the minimal-cost path between $v$ and $t$ must contain at least one vertex $s_i \in S_i$. By definition:

$$c(v, a_i) \leq c(v, s_i) \quad \text{and} \quad c(t, b_i) \leq c(t, s_i)$$

By the subpath property of the minimal-cost path

$$c(v, t) = c(v, s_i) + c(t, s_i) \geq c(v, a_i) + c(t, b_i) = C(v, a_i) - C(t, b_i) = C(v, S_i) - C(t, S_i)$$

Since $c(v, t)$ is always positive, we can, without loss of generality, flip the signs of $C$ so that:

$$c(v, t) \geq |C(v, S_i) - C(t, S_i)|$$

Since this is true for all $S_i$, we have

$$c(v, t) \geq \max_i |C(v, S_i) - C(t, S_i)| = h(v, t)$$

**QED**

Note that the separator itself may not be connected, and even if it is, it may separate the graph into more than two connected components, as in Fig. 3. This does not change any of the arguments above.

### 4.1 Computing the SH Heuristic

Although computing the heuristic is done in a preprocessing stage, it is still important that it be computable somewhat efficiently. In many applications (e.g. traffic-sensitive navigation) the edge weights are dynamic, namely change over time, so the heuristic must be updated periodically to reflect the new weights. Hence efficiency is important.

At first glance, it seems that computing the SH heuristic is much more complex than computing DH. DH requires a single-source minimal-cost computation over the entire graph for each of the $k$ landmarks, costing $O(k(m + n \log n))$ time. Using the same logic, it would seem that computing SH requires similar computation for each vertex in the separators, whose size in a planar graph is $O(\sqrt{n})$ [8], thus costing $O(k\sqrt{n}(m + n \log n))$ time, which is significantly more than the complexity of computing DH. Fortunately, a straightforward "trick" reduces this complexity back down to the same proportions as DH. For each separator $S$, introduce a new "virtual" vertex $w_S$ to the graph connected to all vertices of $S$, and assign a zero weight to all these edges. Then computing the heuristic associated with $S$ is easily seen to be reduced to computing a single-source minimal-cost path computation over the entire new graph for $w_S$.

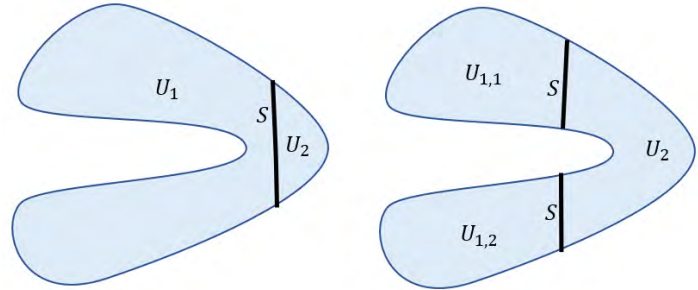

**Figure 3:** A separator may disconnect the graph into **(left)** two or **(right)** more connected components.

### 4.2  Choosing Good Separators

The quality of the SH heuristic very much depends on the choice of separators. It seems that the most informed value of $h(v,t)$ is obtained when $v$ and $t$ are separated by one of the $S_i$ and the separator is compact, in the sense that it contains few vertices *and* these vertices are "close" to each other, i.e. the "cost diameter" of the separator is small. In this case the separator functions as a "bottleneck", through which the minimal-cost path between $v$ and $t$ must pass, and the three points $a_i, b_i$ and $s_i$ mentioned in the proof of Theorem 2 are very close to each other. Indeed, by the triangle inequality

$$c(v,t) \leq c(v,a_i) + c(a_i,b_i) + c(t,b_i) = h(v,t) + c(a_i,b_i)$$

so

$$c(v,t) - h(v,t) \leq c(a_i,b_i)$$

and if $a_i$ and $b_i$ are connected by a path of small cost, $c(a_i,b_i)$ is probably small.

If the separator is not very compact it is difficult to guarantee that $c(a_i,b_i)$ will always be small. Indeed, it is easy to construct simple examples where $c(a_i,b_i)$ is very large. In analogy to the Euclidean planar case, a good rule of thumb is that if the separator is more or less parallel to the *bisector* between $v$ and $t$, $c(a_i,b_i)$ will be small. See Fig. 2 (right). If the graph is a road network that contains highways with small travel times, it is quite effective to use these highways as separators, as they are typically also minimal-cost paths, so all the vertices of the separators are very "close" to each other. Care must be exercised to completely separate the graph along the highway, as typically there are overpasses and underpasses related to the highway, i.e. the graph may not be planar close to the highway.

When the highways do not cover the road network in a systematic manner, it is more practical to take advantage of the planar layout of the network and simply "slice up" the network by straight lines. The simplest approach is to use equally-spaced horizontal and vertical lines. Each such line defines a separator as the vertices on the set of edges intersecting the line, on the one side of the line. However, based on the analogy to planar bisectors mentioned above, it is also advantageous that these lines span a variety of angles. It may also be more practical to use a piecewise-linear polyline to manually (i.e. interactively) define the separator, as this allows to better adapt to the features of the network.

Another way to obtain compact separators is by using the very effective METIS [12] software package for computing compact balanced separators in graphs. See Fig. 4 for example of a polyline separator and one generated by METIS. While we found that METIS generates very compact separators, it completely ignores the "cost diameter" of the separator, so is not optimal for our purposes. We also found that it is difficult to control METIS and cause it to generate a variety of separators at different locations and angles.

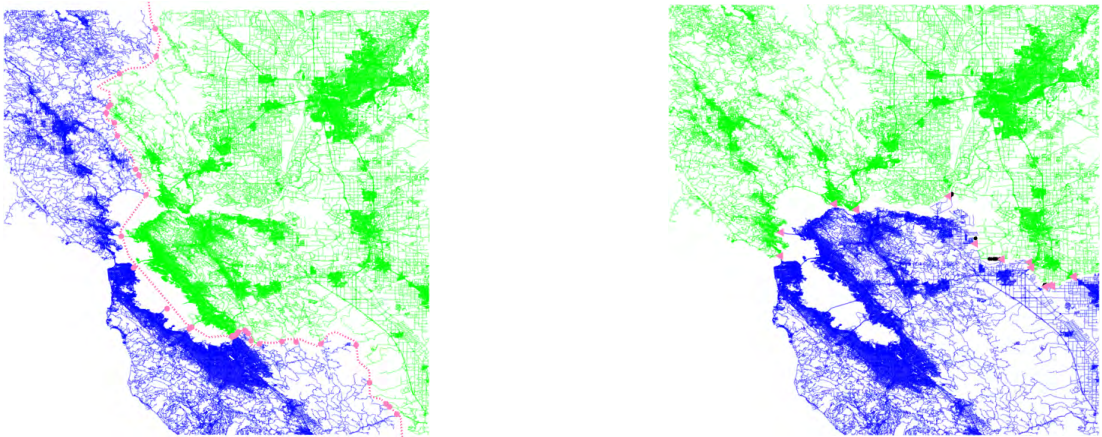

**Figure 4:** Example of separators in the Bay area. **(left)** A separator defined by a dotted pink polyline effectively cutting through bridges across the bay and mountain ridges. The separator is the set of pink vertices which are incident on the cut edges to the left of the polyline. The blue and green regions are the largest connected components in the network after the separator is removed. Black regions are the union of the other (usually very small) connected components. **(right)** A more compact and more balanced separator computed by METIS.

### 4.3  Directed Graphs

The preceding discussion is valid for directed graphs, in which the minimal-cost function is symmetric (thus also a metric): $c(u,v) = c(v,u)$. In reality, road networks are directed graphs, traffic flowing with different velocities in opposite directions, with one-way roads as an extreme case of zero flow in one direction - a fact that cannot be ignored in a real-world application. Thus $c(u,v)$ — the travel time from $u$ to $v$ - will typically be different from $c(v,u)$.

The DH and SH heuristics described above may be generalized to the directed case, by storing *two* values per coordinate, representing minimal costs in opposite directions. For example, given a landmark vertex $l$, the directed triangle inequalities relating to $c(s,t)$ are (see Fig. 5):

$$c(s,l) \leq c(s,t) + c(t,l), \quad c(l,t) \leq c(l,s) + c(s,t)$$

implying:

$$c(s,t) \geq c(s,l) - c(t,l), \quad c(s,t) \geq c(l,t) - c(l,s)$$

Thus the analog to (1) for the DH heuristic in the directed case is:

$$h(s,t) = \max\{c(s,l) - c(t,l), c(l,t) - c(l,s)\} \leq d(s,t)$$

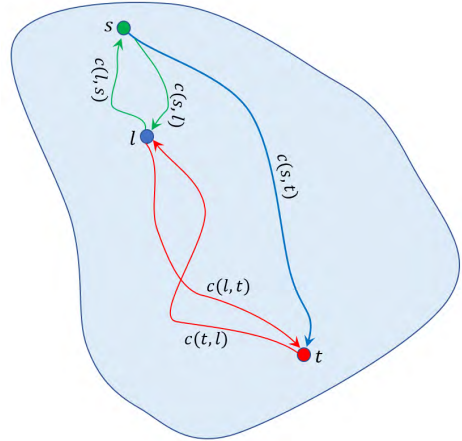

**Figure 5:** Computation of the DH heuristic based on the landmark $l$ in the directed case. The costs of the minimal-cost paths from $l$ in *both* directions are stored for all vertices.

Note that, as opposed to the undirected case, $h(s,t)$ may (in rare cases) be negative, so it should be capped at zero:

$$h(s,t) = \max\{c(s,l) - c(t,l), c(l,t) - c(l,s), 0\}$$

As in the undirected case, using the SH in the directed case requires using separators. These are defined and generated in the same way as the undirected case, i.e. the separation property ignores the directionality of the edges. Since the cost is no longer symmetric, it is not as useful to use the concept of "cost field", rather just the concept of (undirected) connected components. Fig. 6 summarizes the computation of our SH heuristic based on a separator $S$ in the directed case:

---

**Preprocessing:**

For each separator $S$, compute:
1. For every graph vertex $v$, $c(v, S)$ – the minimal cost *from $v$ to $S$*.
2. For every graph vertex $v$, $c(S, v)$ – the minimal cost *from $S$ to $v$*.
3. For every graph vertex $v$, the label of the (undirected) connected component it belongs to when all edges connecting $S$ to other vertices in the graph (in both directions) are removed.

**Online Query:**

Given vertices $u, v$, the SH heuristic based on separator $S$, is:
1. If $u, v$ are in the same connected component, then:
$$h_S(u,v) = \max\{0, c(u,S) - c(v,S), c(S,v) - c(S,u)\}$$
2. If $u, v$ are in different connected components, then
$$h_S(u,v) = c(u,S) + c(S,v)$$

---

**Figure 6:** SH heuristic $h_S(u,v)$ for a directed weight graph based on a separator $S$.

Computing $c(S,t)$ for all vertices $t$ is easy, as discussed in Section 4.1, through the use of a virtual vertex connected with edges of weight zero to all vertices of $S$, and then performing a one-to-all minimal-cost computation from that vertex. At first glance, it would seem that directly computing the opposite $c(t, S)$ is not that straightforward, but this may be solved by reversing the directions of all the graph edges.

## 5. Experimental Results

We have implemented the heuristics mentioned in this paper, namely the differential heuristic (DH), FastMap (FM) and our separator heuristic (SH) and compared how informed they are when approximating the travel time on a number of road networks whose edges are weighted with realistic travel times. We were not able to use the popular benchmark road networks from the 9$^{th}$ DIMACS Implementation Challenge – Shortest Paths dataset [10], because these are *undirected* graphs, so do not reflect reality. Instead, we extracted directed graphs on the equivalent areas of New York, Colorado and the Bay Area from OpenStreetMap [11]. Table 1 shows the specs of those graphs. We were surprised to discover that these were 10x more detailed than those in the DIMACS Challenge. The edges of the graphs were weighted by the minimal travel time along that edge, which was computed as the Euclidean length of the edge (as computed from the latitude and longitude information per vertex) divided by the maximal speed on that edge, as extracted from OpenStreetMap. In our experiments, we randomly chose 10,000 pairs of vertices from each map by randomly choosing two points $(s, t)$ uniformly distributed within the bounding box of the map, and then "snapping" those two points to the closest map vertex, as long as the snap was not too far. We then compared the true fastest path time $c(s, t)$ with the heuristic $h(s, t)$, when varying the number of "coordinates" used in the heuristics between 4, 6 and 8. We performed this experiment for the directed graph and an undirected version of the same graph, where the weight of an edge was taken as the minimal weight of the edges in each direction. In the directed case, we compared SH only to DH, as it is unclear how to generalize FM to the directed case. The DH landmarks were spread uniformly around the boundary of the network. The FM landmark pairs were computed in the manner described by Cohen el al [3], as pairs with distant travel times between them. The SH separators were chosen as a mix of METIS separators and polyline separators specified interactively to take advantage of bottlenecks in the networks.

For each pair of vertices $(s, t)$, we measure the relative *quality* of the heuristic:

$$\text{qual}(s, t) = \frac{h(u, v)}{c(u, v)}$$

which is a value in $[0,1]$ reflecting how informed the heuristic is.

Tables 2 and 3 show the mean and standard deviations of the heuristic qualities for the experiments we performed, on the undirected and directed graphs, respectively. Good values should be between 80% and 100%. The results show that in the undirected case, the SH heuristic is consistently more informed by 3% to 13% than the DH heuristic, which in turn is also 3% to 13% more informed than the FM heuristic. The results are similar in the directed case: SH is 3% to 12% to more informed than DH.

| Graph | New York (NY) | Colorado (COL) | Bay Area (BAY) |
|---|---|---|---|
| **Vertices** | 1,579,003 | 5,154,659 | 3,092,249 |
| **Undirected Edges** | 1,744,284 | 5,400,186 | 3,351,919 |
| **Directed Edges** | 3,104,365 | 10,454,829 | 6,279,871 |

**Table 1:** Statistics of the graphs used in our experiments, as extracted from OpenStreetMap.

| Graph | New York (NY) | | | Colorado (COL) | | | Bay Area (BAY) | | |
|---|---|---|---|---|---|---|---|---|---|
| $k$ | SH | DH | FM | SH | DH | FM | SH | DH | FM |
| 4 | 89 ± 13 | 84 ± 14 | 82 ± 17 | 87 ± 14 | 84 ± 13 | 71 ± 22 | 90 ± 15 | 77 ± 15 | 67 ± 21 |
| 6 | 91 ± 11 | 85 ± 13 | 84 ± 16 | 90 ± 12 | 85 ± 12 | 73 ± 21 | 91 ± 13 | 80 ± 15 | 71 ± 20 |
| 8 | 92 ± 10 | 87 ± 12 | 84 ± 16 | 92 ± 11 | 86 ± 12 | 75 ± 21 | 93 ± 10 | 83 ± 13 | 72 ± 20 |

**Table 2:** Mean and standard deviation of heuristic quality (%) as measured in our experiments, over 10,000 pairs of vertices on an *undirected* road network.

| Graph | New York (NY) | | Colorado (COL) | | Bay Area (BAY) | |
|---|---|---|---|---|---|---|
| $k$ | SH | DH | SH | DH | SH | DH |
| 4 | 88 ± 14 | 83 ± 15 | 87 ± 14 | 84 ± 13 | 89 ± 15 | 77 ± 15 |
| 6 | 91 ± 11 | 85 ± 13 | 90 ± 12 | 85 ± 12 | 91 ± 13 | 80 ± 15 |
| 8 | 92 ± 10 | 87 ± 12 | 92 ± 11 | 86 ± 12 | 93 ± 11 | 83 ± 13 |

**Table 3:** Mean and standard deviation of heuristic quality (%) as measured in our experiments, over 10,000 pairs of vertices on a *directed* road network.

Although a 3% improvement in the quality of the SH heuristic over the DH heuristic would seem rather small, it can make a surprisingly big difference in the performance of the A* algorithm. The effect of a good heuristic is to reduce the number of road network vertices traversed during the search for the fastest path. Thus the *efficiency* of a heuristic in conjunction with A* is measured as the number of vertices on the fastest path divided by the total number of vertices traversed by A*:

$$\text{eff}(s,t) = \frac{\#vertices(\text{fastest\_path}(s,t))}{\#vertices(\text{A*\_traversal}(s,t))}$$

The closer this number is to 1 – the more efficient the heuristic is. The efficiency of the heuristic is the mean of this quantity over all possible pairs $(s,t)$. The best possible efficiency on a road network is typically 40%-50%, since any variant of A* must traverse at least the fastest path vertices and also their immediate neighbors. When a heuristic is used, the efficiency can drop dramatically to the vicinity of 1%, meaning 100 vertices of the graph are explored for every one vertex along the fastest path. Tables 4 and 5 compare the efficiencies of the different heuristics using the same formats as Tables 2 and 3. SH is more efficient than DH by a factor between 1.35 and 2.4 in the undirected case, and between 1.26 and 2.67 in the directed case.

Figs. 7 and 8 give more details of the results for the simplest case of $k = 4$ on undirected and directed road networks. The left column of each table illustrates the four DH landmarks in red, the four FM pairs in blue and the four SH polyline separators in four other colors. The middle column shows the histogram of the distribution of the qualities of DH, FM and SH values in red, blue and green, respectively. The right column shows the histogram of the efficiencies, color-coded in the same way.

| Graph | New York (NY) | | | Colorado (COL) | | | Bay Area (BAY) | | |
|---|---|---|---|---|---|---|---|---|---|
| $k$ | SH | DH | FM | SH | DH | FM | SH | DH | FM |
| 4 | 6.0 ± 9.9 | 3.6 ± 9.4 | 3.2 ± 6.7 | 3.4 ± 5.9 | 2.9 ± 6.4 | 1.5 ± 3.1 | 4.8 ± 8.9 | 3.5 ± 10.4 | 2.0 ± 5.1 |
| 6 | 7.1 ± 11.0 | 5.7 ± 11.9 | 3.7 ± 7.4 | 5.8 ± 8.5 | 3.8 ± 8.6 | 1.9 ± 4.9 | 8.1 ± 15.3 | 3.9 ± 10.8 | 2.7 ± 6.5 |
| 8 | 8.1 ± 12.5 | 6.0 ± 12.0 | 3.7 ± 7.5 | 7.3 ± 9.3 | 4.4 ± 9.7 | 2.2 ± 5.5 | 11.3 ± 17.7 | 4.7 ± 11.5 | 3.0 ± 7.1 |

**Table 4:** Mean and standard deviation of heuristic efficiency (%) for A* as measured in our experiments, over 1,000 pairs of vertices on an *undirected* road network.

| Graph | New York (NY) | | Colorado (COL) | | Bay Area (BAY) | |
|---|---|---|---|---|---|---|
| $k$ | SH | DH | SH | DH | SH | DH |
| 4 | 5.6 ± 9.8 | 3.1 ± 7.9 | 3.4 ± 6.0 | 2.7 ± 5.6 | 4.8 ± 9.5 | 3.3 ± 10.8 |
| 6 | 6.7 ± 11.3 | 6.0 ± 12.3 | 5.8 ± 8.0 | 3.6 ± 8.2 | 8.0 ± 15.3 | 3.6 ± 11.2 |
| 8 | 8.2 ± 13.2 | 6.1 ± 12.3 | 7.4 ± 9.0 | 4.2 ± 9.4 | 11.2 ± 18.3 | 4.2 ± 11.9 |

**Table 5:** Mean and standard deviation of heuristic efficiency (%) for A* as measured in our experiments, over 1,000 pairs of vertices on a *directed* road network.

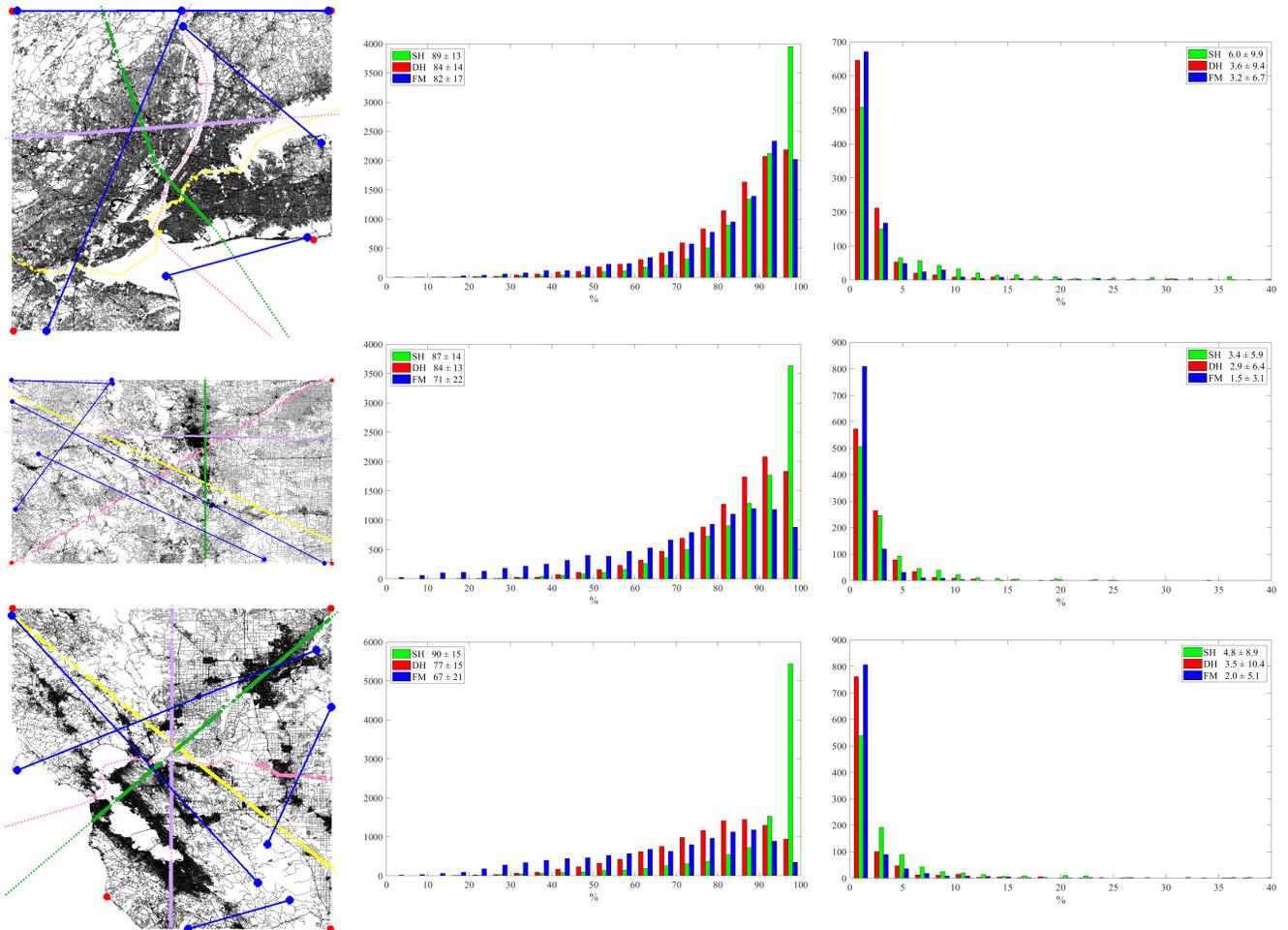

**Figure 7:** Comparison of heuristics using $k = 4$ coordinates on *undirected* weighted road networks. Red points are DH landmarks. Blue points joined by line segments are FM pairs. SH separators are in other colors. **Top:** New York (NY). **Middle:** Colorado (COL), **Bottom:** Bay Area (BAY). **Left:** Road network, **Middle:** Heuristic quality histogram, **Right:** Heuristic efficiency histogram.

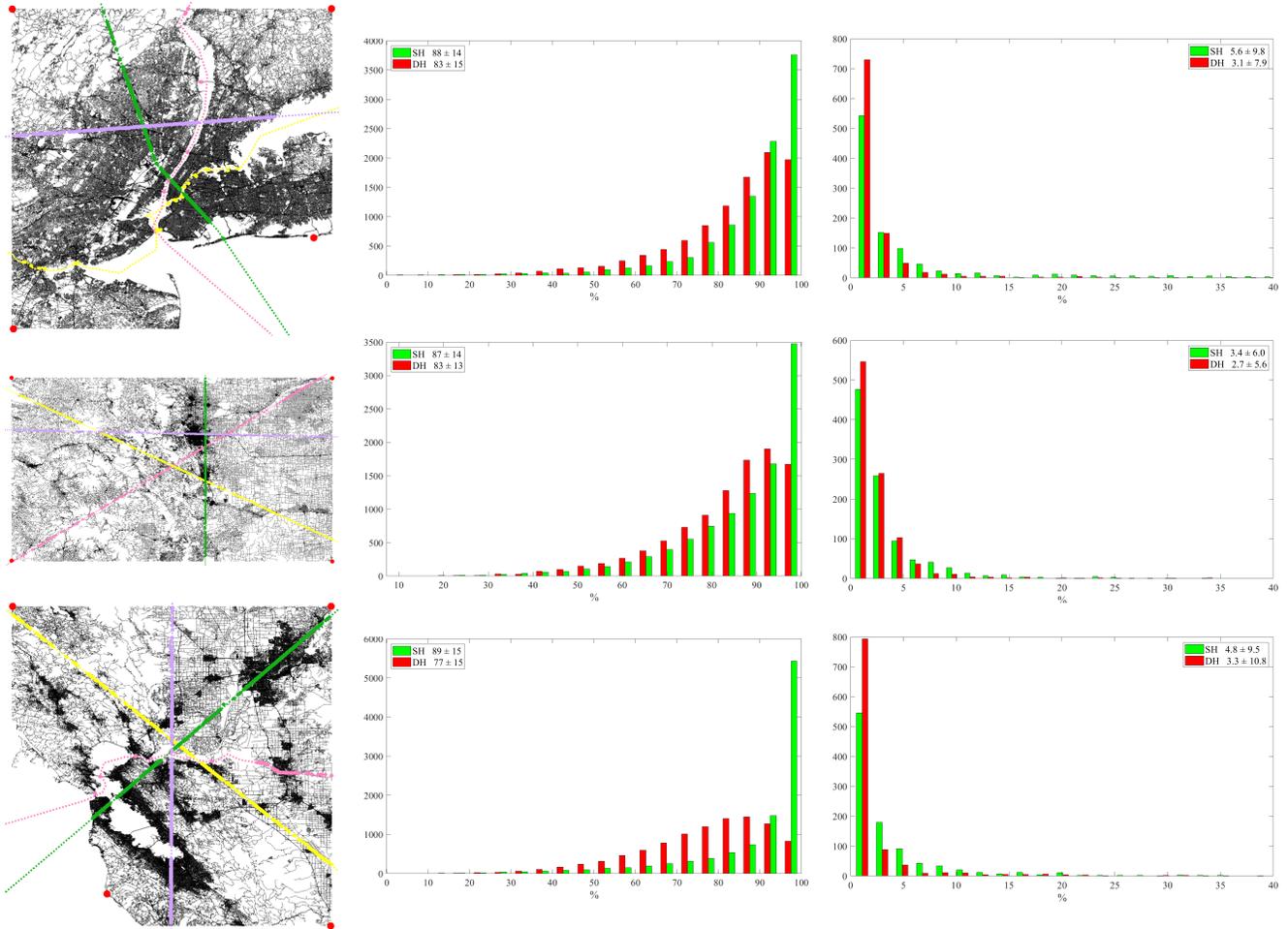

**Figure 8:** Comparison of heuristics using $k = 4$ coordinates on *directed* weighted road networks. Red points are DH landmarks. Blue points joined by line segments are FM pairs. SH separators are in other colors. **Top:** New York (NY). **Middle:** Colorado (COL), **Bottom:** Bay Area (BAY). **Left:** Road network, **Middle:** Heuristic quality histogram, **Right:** Heuristic efficiency histogram.

To illustrate better the efficiency of the SH heuristic compared to that of the DH heuristic, Figs. 9 and 10 show the vertices traversed by A* when searching for the fastest path using the different heuristics on the same $(s, t)$ pair, in the undirected and directed cases. Despite the modest improvements in quality between DH and SH, the efficiency is improved by anywhere between a factor of 2.9 and a factor of 30.

It is interesting to understand better the effect of the location of the separator on the efficiency of A* using SH. Fig. 11 shows how A* traverses a road network when searching for the fastest path between two vertices using SH based on a single separator, in three different locations. For simplicity, this network is undirected and its edges are weighted by Euclidean edge lengths. The separators are all parallel to the "bisector" between the two vertices, but at different distances from the target. As long as the vertex under investigation is separated from the target vertex, the heuristic seems to be quite informed. This changes, sometimes quite dramatically, when the separator is crossed, indicating that the true power of the heuristic is in its separation property, as opposed to, e.g. the DH heuristic, which is based on no more than the very basic triangle inequality.

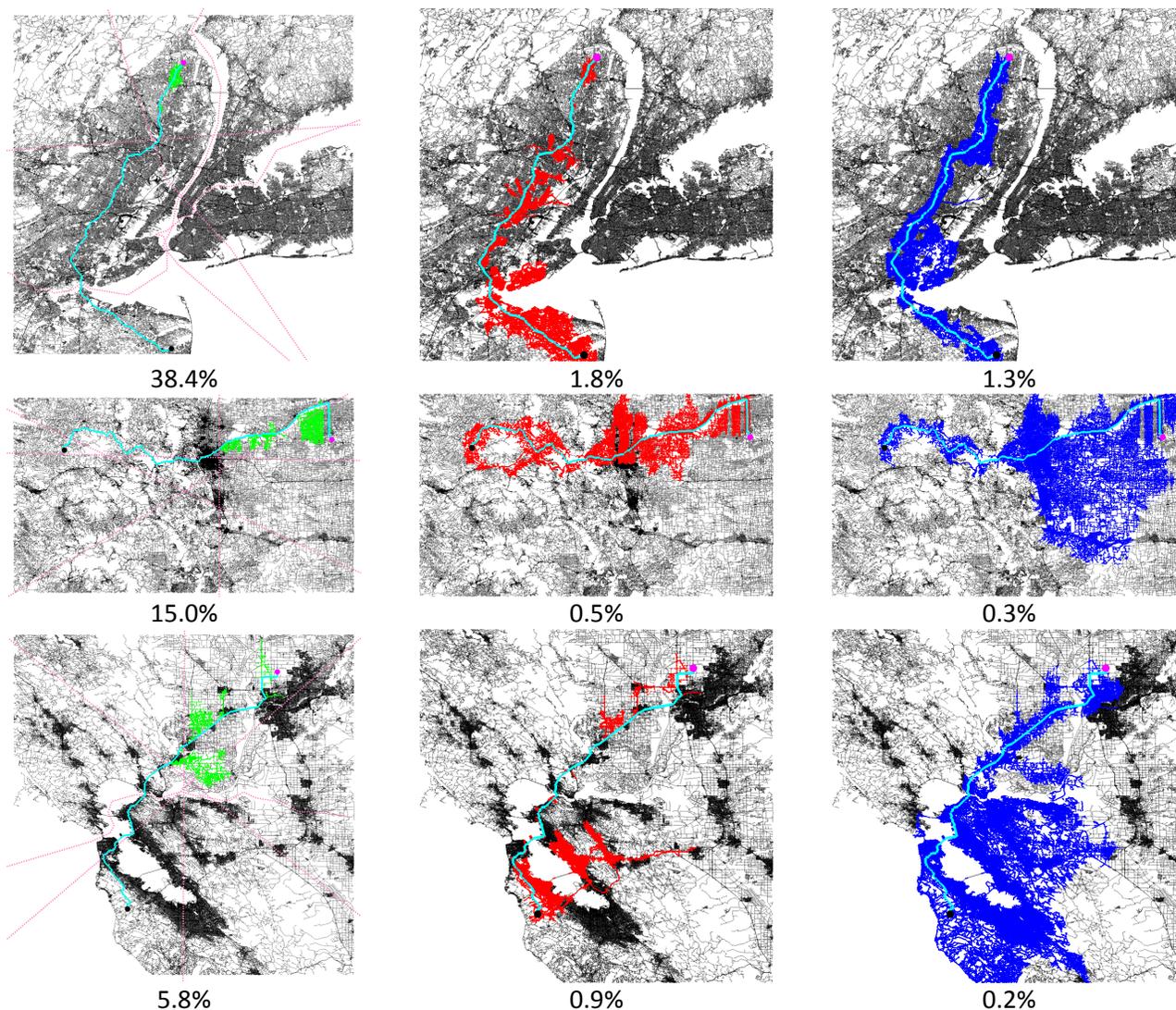

**Figure 9:** Efficiency of the heuristics with $k=4$ coordinates on different weighted *undirected* road networks. **Top:** NY, **Middle:** COL, **Bottom:** BAY. Colored vertices show the vertices traversed during the A* search for the fastest path from the black vertex to the magenta vertex. Green – SH, Red – DH, Blue – FM. Magenta dotted lines indicate the SH separators. Cyan indicates fastest path, usually taking advantage of highways.

## Summary and Conclusion

We have proposed a relatively simple way to compute an admissible and consistent heuristic SH for the A* algorithm for computing the minimal-cost path in a weighted directed graph. In some sense, this heuristic may be viewed as a powerful generalization of the differential heuristic DH (originally called ALT), which has proven to be very effective in its own right. SH is based on the notion of graph separators, which may be generated automatically or manually on road networks, and is shown experimentally to be of higher quality (i.e. more informed) than DH by about 10%, but resulting in an increase in efficiency of up to an order of magnitude, when used by A* to generate fastest paths in directed road networks with edges weighted by travel times.

SH is applicable to both undirected and directed graphs and seems to perform similarly on both.

Like DH, SH may be used in conjunction with other types of optimizations of the A* algorithm (e.g. bi-directional search, reach-based and hierarchical methods) to independently boost its performance.

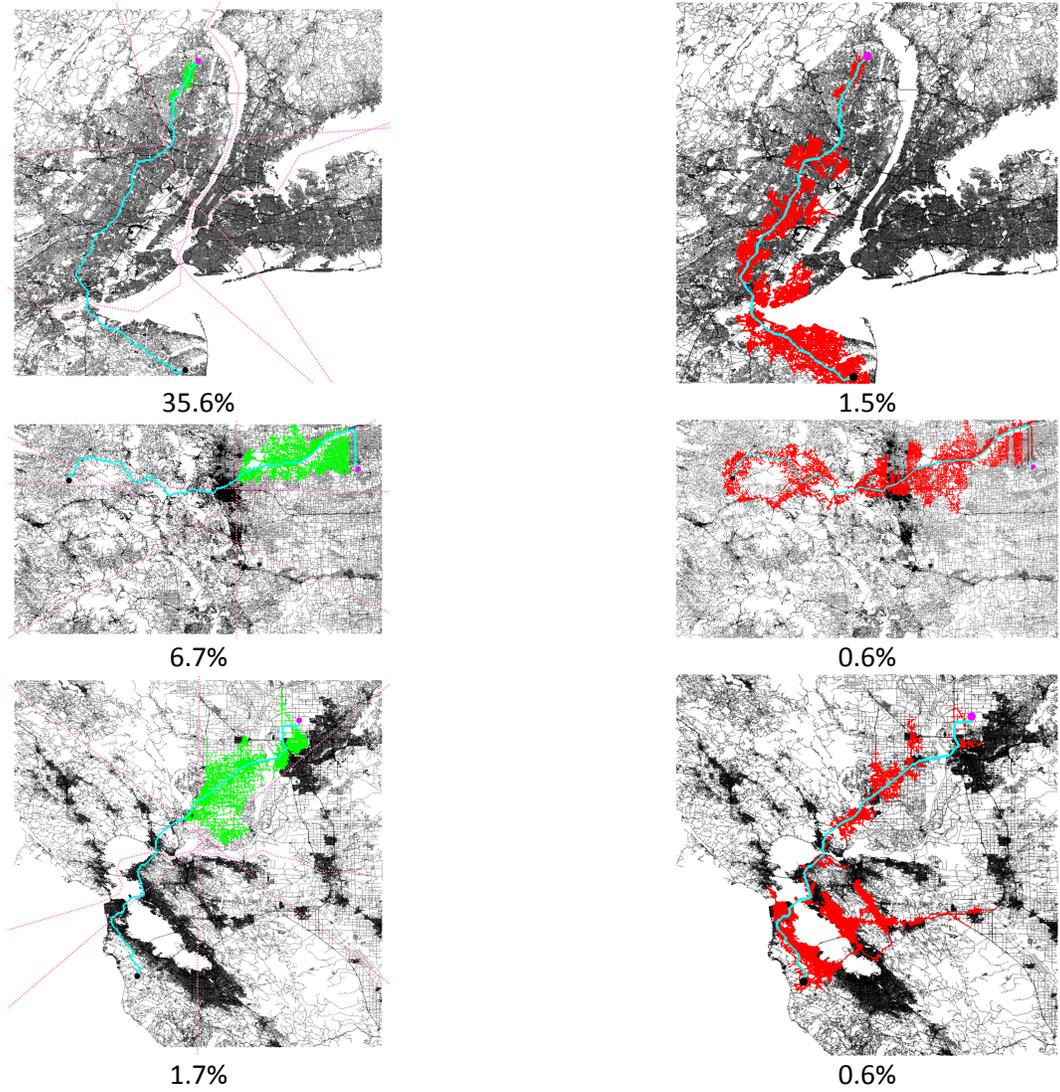

**Figure 10:** Efficiency of the heuristics with $k = 4$ coordinates on different *directed* road networks. **Top:** NY, **Middle:** COL, **Bottom:** BAY. Colored vertices show the vertices traversed during the A* search for the fastest path from the black vertex to the magenta vertex. Green – SH, Red – DH. Magenta dotted lines indicate the SH separators. Cyan indicates fastest path, usually taking advantage of highways.

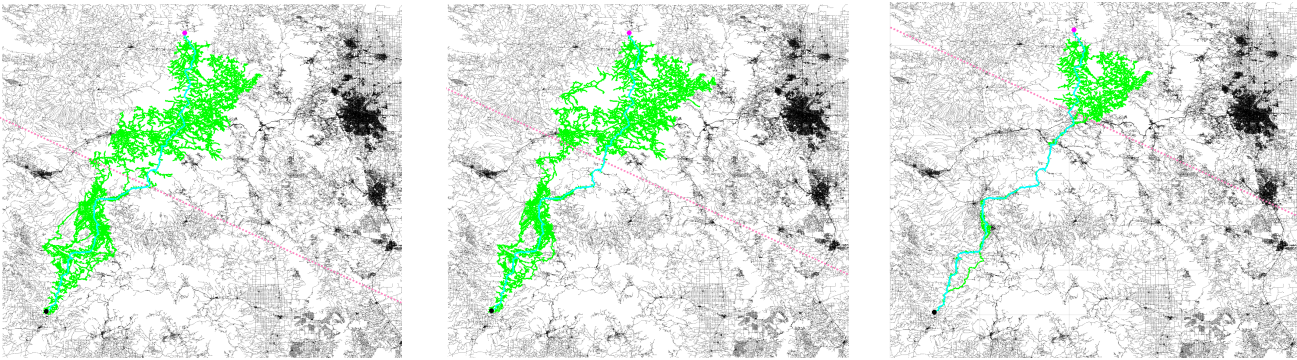

**Figure 11:** The effect of the location of the separator on the efficiency of the SH heuristic in an undirected road network whose edges are weighted by Euclidean edge lengths. Green vertices are those traversed by A* using SH with a single separator, marked in magenta, when computing the fastest path from the black source to the magenta target vertex. The separator is parallel to the bisector between the two vertices, but at different distances from the target. Note the deterioration in the efficiency once the separator is crossed.